\documentclass[10 pt,final,journal,letterpaper,oneside,onecolumn]{IEEEtran}
\usepackage{graphicx}
\usepackage{amssymb,amsmath,amsfonts}
\usepackage{suffix}
\usepackage{amssymb}
\usepackage{mathtools}
\usepackage{cite}  
\usepackage{amsthm}
\usepackage[utf8]{inputenc}
\usepackage[english]{babel}
\usepackage{epstopdf}
\usepackage{color}
\usepackage{times}
\usepackage{fancyhdr,graphicx,amsmath,amssymb}
\usepackage{algorithmicx}
\usepackage[ruled]{algorithm}
\usepackage{algpseudocode}
\ifCLASSINFOpdf
\else
\fi

\begin{document}
%
\title{Efficient Mining Cluster Selection for Blockchain-based Cellular V2X Communications}

\author{Furqan Jameel, Muhammad Awais Javed, Sherali Zeadally, and Riku J\"antti \thanks{Furqan Jameel and Riku J\"antti are with the Department of Communications and Networking, Aalto University, 02150 Espoo, Finland. (email: furqanjameel01@gmail.com and riku.jantti@aalto.fi).

Muhammad Awais Javed is with Electrical and Computer Engineering Department, COMSATS University Islamabad, Islamabad, Pakistan (email: awais.javed@comsats.edu.pk)

Sherali Zeadally is with the College of Communication and Information, University of Kentucky, Lexington, KY 40506-0224, USA (email: szeadally@uky.edu).
}}%
\markboth{ACCEPTED IN IEEE TRANSACTIONS ON INTELLIGENT TRANSPORTATION SYSTEMS}%
{\MakeLowercase{\textit{et al.}}: ACCEPTED IN IEEE TRANSACTIONS ON INTELLIGENT TRANSPORTATION SYSTEMS}

\maketitle

\begin{abstract}
Cellular vehicle-to-everything (V2X) communication is expected to herald the age of autonomous vehicles in the coming years. With the integration of blockchain in such networks, information of all granularity levels, from complete blocks to individual transactions, would be accessible to vehicles at any time. Specifically, the blockchain technology is expected to improve the security, immutability, and decentralization of cellular V2X communication through smart contract and distributed ledgers. Although blockchain-based cellular V2X networks hold promise, many challenges need to be addressed to enable the future interoperability and accessibility of such large-scale platforms. One such challenge is the offloading of mining tasks in cellular V2X networks. While transportation authorities may try to balance the network mining load, the vehicles may select the nearest mining clusters to offload a task. This may cause congestion and disproportionate use of vehicular network resources. To address this issue, we propose a game-theoretic approach for balancing the load at mining clusters while maintaining fairness among offloading vehicles. Keeping in mind the low-latency requirements of vehicles, we consider a finite channel blocklength transmission which is more practical compared to the use of infinite blocklength codes. The simulation results obtained with our proposed offloading framework show improved performance over the conventional nearest mining cluster selection technique.      
\end{abstract}


\IEEEpeerreviewmaketitle

\section{Introduction}

Blockchain is a technology that leverages the subtle interaction of multiple data structures and incentive mechanisms. In isolation, the various components that comprise the blockchain are well known and in some cases have existed for years \cite{xiao2020survey}. The novelty of blockchain stems from a combination of these elements that could be applied to different applications ranging from cryptocurrencies to smart contracts, and from resource management to differential privacy. The rapid success of blockchain has generated a lot of interest in the design principles employed in the development of large-scale systems. This, in turn, has prompted researchers in academia and industry to critically reassess traditional methods used to process information \cite{8977442}. The goal is to determine the extent to which the architectural aspects of blockchain might be replicated in analogous scenarios to reduce or eliminate current inefficiencies.

The integration of blockchain with cellular vehicle-to-everything (V2X) communications is expected to revolutionize intelligent transportation systems (ITSs), in terms of enriched travel experiences, on-road safety services, and transportation efficiency \cite{storck20195g}. Generally speaking, the V2X communication enabled by a cellular infrastructure is known as cellular V2X and encompasses the inter-vehicle (V2V) communications, vehicle-to-pedestrian (V2P) communications, and vehicle-to-infrastructure (V2I) communications \cite{jameel2018propagation}. \textcolor{black}{The efficient architecture of blockchain and the cooperative/interactive nature of cellular V2X communications can solve various resource-related issues. Blockchain-based cellular V2X communications can also overcome the conventional resource-constrained barriers (such as security and privacy costs) and can improve real-time exchange of information, spectrum allocation, and mobile offloading. Recent advances in blockchain-based vehicular communications have opened up opportunities in various areas of vehicular communications. This includes IEEE 802.11p based vehicular networks where blockchain is used for authentication and key management \cite{lei2017blockchain} and cellular V2X networks for guaranteeing security \cite{muthanna2019secure,zhou2019secure}. Some other important areas include efficient and transparent energy trading between electric vehicles and grid stations, and reliable platoon management for autonomous vehicles \cite{hussain2018autonomous}. Additionally, blockchain-enabled vehicles can act as secure and trustworthy data collection platforms for transportation authorities.} 

\textcolor{black}{Mining is an important process in blockchain technology for securing the validity and integrity of transactions \cite{jameel2020bloc}. There are two types of mining schemes, i.e., solo mining and pool/cluster mining \cite{wang2015exploring}. As the name suggests, cluster mining involves multiple miners to mine a block, whereas, solo mining is performed individually. Regardless of the mining approach, mining a block is a computation-intensive and resource-demanding task, especially for wireless vehicular networks. To improve mining efficiency, the mining task can be offloaded to nearby vehicles. This work proposes a mining cluster selection strategy for efficiently completing the mining task while maintaining fairness among mining vehicles.} 

\subsection{Related Work}

Vehicular communication is a significantly mature concept and has been extensively explored in the literature. Performance analysis and the evaluation of vehicular communications were some of the key areas of interest for researchers initially. Some of these studies focused on the deterministic evaluation of performance while others focused on the wireless channel characterization for inter-vehicle communications \cite{8985250,khan2020multiobjective,jameel2019performance,jeyaraj2017reliability}. However, these studies lack the fundamentals of exchanging data between the vehicle and the infrastructure and they used relatively simplistic spatial modeling techniques. To extend these studies, Guha \emph{et al.} first performed the analysis for cellular V2X communication \cite{guha2016cellular}. Specifically, they analyzed the association probabilities of the vehicle and proposed a maximum-power association technique to enhance network performance. From the perspective of resource optimization, some studies on cellular V2X communications also exist where the authors considered a unicast communication paradigm for guaranteeing the sum-rate for V2X communication \cite{masmoudi2019efficient}. Similarly, the authors of \cite{mei2018latency} proposed a resource allocation solution to maximize the service rate of the network users while Masmoudi \emph{et al.} reduces the latency of V2X communication to support driver safety applications \cite{masmoudi2018mixed}. 
 
Although there many studies on blockchain-based vehicular communications \cite{shrestha2019new,ali2020integrating}, only a few works focus on the mining task offloading aspects of such networks. For instance, Sharma \emph{et al.} in \cite{sharma2018energy} focused on energy-efficient offloading of data for Internet-of-vehicles (IoV). They showed that their proposed model optimally controls the number of transactions in a distributed manner. Similarly, the authors of \cite{zhou2019secure,zhou2018blockchain} proposed an energy-efficient trading solution for vehicle-to-grid (V2G) communications. They developed an iterative convex-concave algorithm to address the social welfare optimization problem in blockchain-based vehicular networks. Here, the social welfare measures the optimal quality of resources allocated in the network. In \cite{de2019energy}, the authors proposed energy and proof-of-work aware offloading mechanism for vehicular ad-hoc networks (VANETs) while the authors of \cite{liu2018adaptive} proposed an adaptive participation scheme for charging the electric vehicles from the smart grid. Other studies \cite{chen2019secure,jindal2019survivor,kang2017enabling} also loosely focus on the task offloading aspect of blockchain-based vehicular communications, but their main focus is still on security and energy efficiency aspect of such networks. \textcolor{black}{The review of these various recent works demonstrates that there are only a few works \cite{chen2019secure,jindal2019survivor,kang2017enabling} on mining task offloading in the literature of blockchain-based vehicular communications. However, the focus of these studies is not on finite blocklength which is not practical for mission-critical applications \cite{wang2019secure} because most practical systems work in finite blocklength regime. As a feasible solution, the finite blocklength approach is being rapidly adopted for enabling ultra-reliable low-latency vehicular communications \cite{yang2020ultra}. In this work, we aim to identify the performance limits of vehicular communications in finite blocklength regime and propose a solution to address the mining task offloading problem for blockchain-enabled V2X communications.}

\subsection{Motivation and Contribution}

Motivated by the above-mentioned studies, we anticipate that the interchange of homogeneous data facilitated by blockchain technologies, is the catalyst needed to transform the ideas of blockchain-enabled autonomous vehicles into practice because of peer-to-peer and distributed nature of blockchain. However, before the full realization of such autonomous vehicles, there is a need to further explore and solve the challenges associated with the blockchain-based cellular V2X communications. One such challenge is the execution of mining tasks and the issues related to offloading such tasks to the mining cluster. The initiatives described in this work are attempts at breaking down the nascent data silos generated by cellular V2X communications in the blockchain ecosystem. In the process, we aim to provide a feasible solution for selecting the appropriate mining cluster for offloading the mining tasks. We summarize our main research contributions as follows: 
\begin{enumerate}
\item  We consider a practical finite blocklength (number of wireless channel uses) transmission architecture for blockchain-based cellular V2X communications. This guarantees low-latency transmission due to the short blocklength. However, it suffers from non-zero decoding error probability. To the best of our knowledge, such finite blocklength transmission has not yet been considered for blockchain-based cellular V2X communications.  
\item We formulate the mining task offloading problem by considering the transmission rate of vehicles and the available computation resources in the mining cluster. Due to the limited local view of the offloading vehicles, the formulated problem ensures the feasible execution of mining tasks.
\item We propose a game-theoretic solution to balance the computation load at the mining cluster. The solution also ensures that all the offloading vehicles offload their task fairly. The simulation results show performance improvements over the conventional nearest cluster selection approach.
\end{enumerate}

\subsection{Organization}

The remainder of the paper is organized as follows. Section II presents the system model and problem formulation. In Section III, we describe in detail the proposed optimization framework. Section IV presents the simulation results and relevant discussions. Finally, Section V makes some concluding remarks and future work. Table \ref{tab1} presents a list of commonly used notations in this work.

\begin{table}
\centering
\caption{Commonly used notations.}
\label{tab1}
\begin{tabular}{|p{1.5cm}|p{6cm}|}
\hline
\textbf{Symbol} & \textbf{Definition} \\ \hline
$\Omega_{i,j}$ & Average transmission rate \\ \hline
$R_{i,j}$ & Computation resources allocated by mining cluster \\ \hline
$\phi_{i,j}$ & Offloading indicator \\ \hline
$S_j$ & Maximum number of vehicles that can be offloaded \\ \hline
$\mathcal{F}$ & Set containing all mining clusters \\ \hline
${\mathcal{G}}$ & Set containing all offloading vehicles \\ \hline
$h_{i,j}$ & Channel gain \\ \hline
$p_j^k$ & Price set by $j$-th mining cluster \\ \hline
$P_i$ & Transmit power of offloading vehicle \\ \hline
$B_{i,j}$ & Bandwidth allocated to offloading vehicle \\ \hline
$m_{i,j}$ & Margin for $i$-th offloading vehicle for offloading to $j$-th mining cluster \\ \hline
$w_{i,j}$ & Weight of the edge between offloading vehicle and mining cluster \\ \hline
$W_c$ & Coherence bandwidth \\ \hline
$c$ & Constant for balancing the utilities in price \\ \hline
$n_{i,j}$ & Number of bandwidth units allocated to offloading vehicle \\ \hline
$Q^{-1}(.)$ & Inverse Q-function \\ \hline
$\sigma^2$ & Variance of noise \\ \hline
\end{tabular}
\end{table}

\section{System Model and Problem Formulation}

In this section, we present the details of the system model along with a discussion on the problem formulation.
\begin{figure*}[!htp]
\centering
\begin{tabular}{c}
\includegraphics[trim={0 0cm 0 0cm},clip,scale=.48]{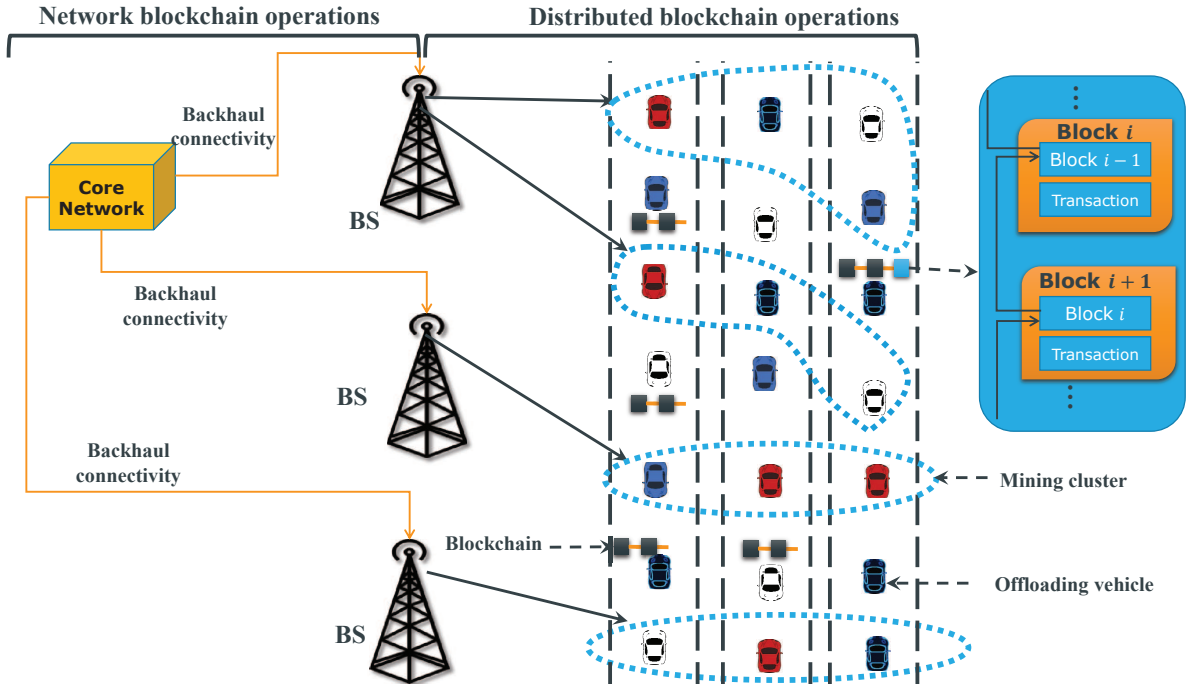}
\end{tabular}
\caption{Blockchain-based cellular V2X communications. The network consists of number of base stations (BSs) connected to the core network and Internet via backhaul links. The BSs serve different vehicles in the coverage region and manage the transaction-messages between vehicles and the core.}
\label{fig1}
\end{figure*}

\subsection{\textcolor{black}{Network setup}}

Fig. \ref{fig1} shows a distributed mining scenario with $NV+M$ vehicles which are considered to form $N$ mining clusters in the set $\mathcal{F}$ wherein each cluster contains $V$ vehicles. The mining clusters of vehicles are assumed to have the computation resources to provide a task execution service to the set $M$ vehicles represented in set ${\mathcal{G}} $. \textcolor{black}{In a practical setting, vehicles can switch between normal and mining modes depending on their computation resources. Once in the mining mode, the vehicle does not change mode until the completion of the mining task. The impact of switching between mining and normal modes and their relevant processes is beyond the scope of this study.} To perform the recording of transactions in the network, all the vehicles are assumed to run blockchain applications. \textcolor{black}{Accordingly, $M$ vehicles with lower computation resources may offload the computationally difficult tasks to one of the mining clusters.}

When a vehicle $i \in {\mathcal{G}} $ uploads a task to the mining cluster $j \in \mathcal{F} $ a task $X$, the head of the mining cluster receives the task and distributes it among cluster members. Due to the delay-sensitive nature of vehicular communications, it is not feasible to allocate different timeslots. Thus, we assume that all vehicles communicate at the same time $T$ over orthogonal frequency bands. The total frequency band is divided into $B_{0}$ bandwidth units, such that the bandwidth allocated to an offloading vehicle communicating to the mining cluster is expressed as $B_{i,j}=n_{i,j}B_{0}$. Here, $n_{i,j}$ denotes the number of bandwidth units allocated to the $i$-th vehicle for communicating to the $j$-th mining cluster. We also assume that the coherence bandwidth $W_{c}=n_{max }B_{0}$ is divisible by $B_{0}$ and that the total bandwidth allocated to the vehicles is not higher than $W_{c}$. Thus, we have: 
\begin{align}
\sum_{i=1}^{M}{\sum_{j=1}^{N}{n_{i,j}}}\le n_{max }
\end{align}

\textcolor{black}{In general, most theoretical results reported in the literature are asymptotic in the number of channel uses (blocklength), i.e., they are valid when the number of channel used approaches infinity \cite{devroye2010information}. However, when the blocklength is very large, the decoding error rate becomes very small \cite{dosti2019performance}. Such an opportunity may not be available in practical vehicular communication conditions due to requirement of low latency and short blocklength. Due to this finite short blocklength, the probability of non-zero decoding error significantly increases which is often neglected in the analysis of blockchain-based vehicular networks.} Assuming the channel between the offloading vehicle and the head of mining cluster follows Rayleigh fading, the maximum upload transmission rate for a given channel blocklength $L_{i,j}=B_{i,j}T$ can be expressed as:
\begin{align}
\Omega_{i,j}=n_{i,j}B_{0}T \Biggl[ \log _{2}\biggl\lbrace 
1+\frac{P_{i}g_{i,j}}{n_{i,j}}\biggl\rbrace 
-\sqrt{\frac{U_{i,j}}{n_{i,j}B_{0}T}}\frac{Q^{-1}(\varpi)}{\ln 
2}\Biggr]
\label{eq_1}
\end{align}
where $P_{i,j}$ is the transmit power for $i$-th vehicle , $g_{i,j}=\frac{\vert h_{i,j}\vert ^{2}}{(B_{0}\sigma ^{2})}$ with $h_{i,j}$ denoting the channel gain and $\sigma ^{2}$ represents the variance of additive white Guassian noise (AWGN). Moreover, $Q^{-1}(.)$ is the inverse Q-function $Q(x)$ given as
\begin{align}
Q(x)=\int_{x}^{\infty }{\frac{1}{\sqrt{2\pi }}e^{-\frac{t^{2}}{2}}}dt,
\end{align}
and $U_{i,j}$ denotes the random variability of the channel when compared to a deterministic channel with the same capacity \cite{ren2019resource} written as
\begin{align}
U_{i,j}=1-\biggl(1+\frac{P_{i}g_{i,j}}{n_{i,j}}\biggr)^{-2}.
\end{align}

\subsection{Problem Formulation}

In this work, our objective is to match the offloading vehicle with the appropriate mining cluster based on the price and computation resources. This is intuitive because different mining clusters may have different resources and, thus, different costs to provide task execution services. In this context, two key factors must be considered to efficiently perform the computation of the offloading task. First, there is the transmission rate for a specific task offloading vehicle. As shown in (\ref{eq_1}), the transmission rate mainly depends on the received signal of the vehicle for decoding the message. Second, task execution also depends on the number of available computation resources in the mining cluster. We denote the portion of computation resources of the $j$-th cluster that is assigned for performing the task of the $i$-th vehicle as $R_{i,j}$. Furthermore, since multiple vehicles can offload a task to the mining cluster, it is important to define an offloading indicator as $\phi_{i,j}$. The offloading indicator identifies whether a particular vehicle is offloading a task to the $j$-th mining cluster such that: 
\begin{align}
\phi_{i,j}=\Biggl \lbrace \begin{matrix}
1 & i\text{-th vehicle offloading to } j \text{-th mining cluster} & \\
0 & \text{otherwise} & \\
\end{matrix}
\end{align}

\begin{figure}[!htp]
\centering
\begin{tabular}{c}
\includegraphics[trim={0 0cm 0 0cm},clip,scale=.32]{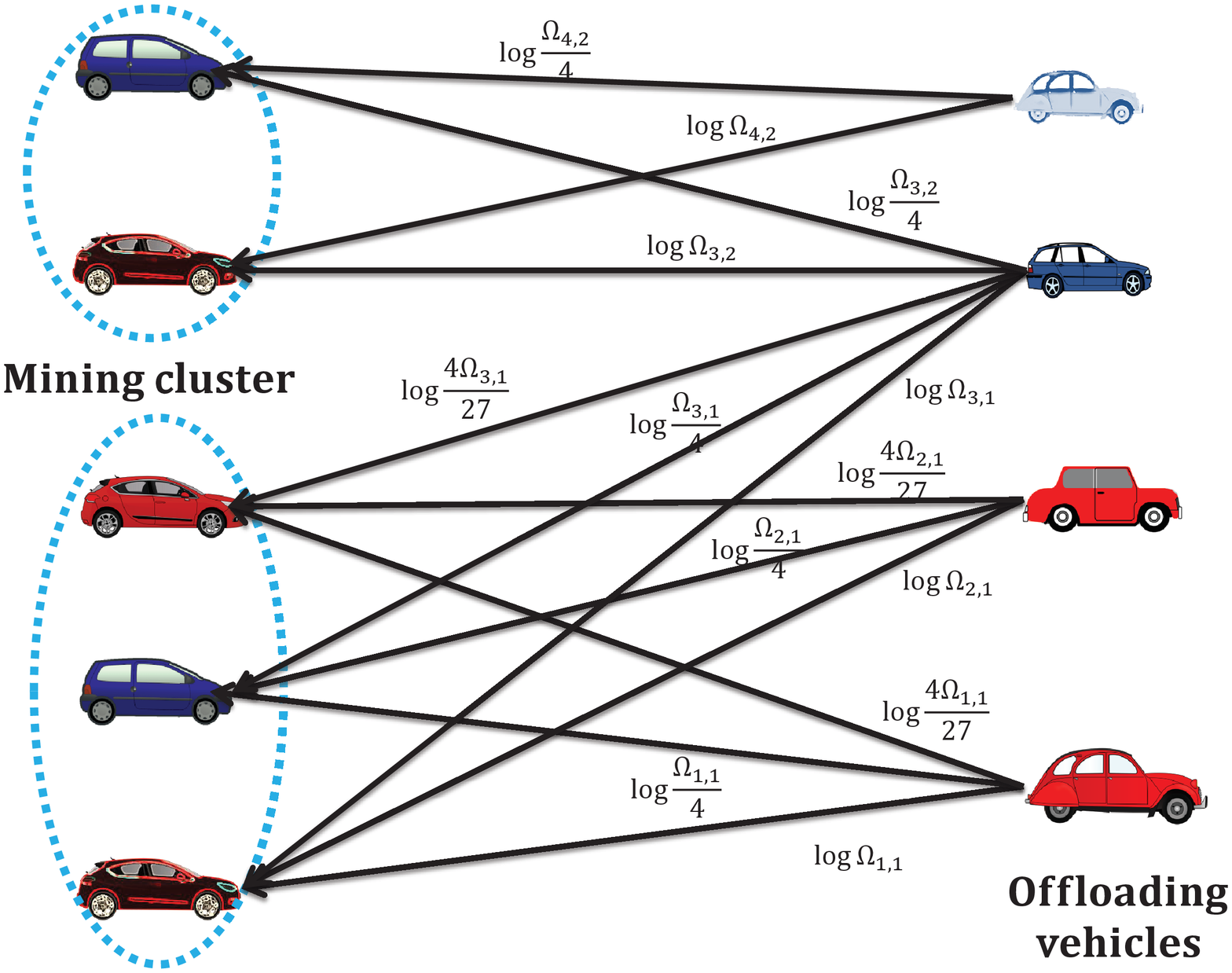}
\end{tabular}
\caption{Graph illustration for selecting appropriate mining cluster.}
\label{fig2}
\end{figure}

Thus, the offloading problem can now be written as: 
\begin{align}
\max{\sum_{i=1}^{M}{\log \left(\sum_{j=1}^{N}{R_{i,j}\Omega_{i,j}\phi_{i,j}}\right)}},
\label{eq_2}
\end{align}
\vspace{-0.3cm}
\begin{align}
\text{s.t.}\ \ \ &\textbf{C1: }\sum_{i=1}^{M}{R_{i,j}}\le 1, \nonumber \\
&\textbf{C2: }\sum_{j=1}^{N}{\phi_{i,j}}\le 1, \nonumber \\
&\textbf{C3: }\phi_{i,j} \in \lbrace 0,1\rbrace, \nonumber \\ 
&\textbf{C4: }R_{i,j}\ge 0, \nonumber 
\end{align}
where the logarithmic function has been used as a common choice of the utility function for maintaining fairness. It is worth noting that the linear utility functions result in a trivial solution for the case of objective maximization, where providing more resources to vehicles with low transmission rates is desirable. Hence, following the approach of \cite{kelly1997charging}, we use a logarithmic utility function in the remainder of this paper which is closer to the resource allocation philosophy of practical wireless networks. Moreover, in (\ref{eq_2}), the expression $\sum_{j=1}^{N}{R_{i,j}\Omega_{i,j}\phi_{i,j}}$ shows the task offloading gain that the $i$-th vehicle can get by offloading the task to the $j$-th mining cluster. Here, the constraint \textbf{C1} indicates that the $j$-the mining cluster does not over-utilize its computational resources while serving different vehicles. The constraint \textbf{C2} ensures that a particular vehicle can only offload the task to a single mining cluster. The constraints \textbf{C3} and \textbf{C4} highlights the bounds of computation resources and the binary nature of the offloading indicator. 

It is can be noted from (\ref{eq_2}) that the objective is to maximize the global offloading utility. In this way, the offloading among the overcrowded mining clusters can be balanced by matching the offloading vehicle with an appropriate mining cluster. Moreover, it is worth mentioning that the problem in (\ref{eq_2}) is a mixed integer programming problem and it is difficult to solve in polynomial time. Though relaxations can be performed on the offloading indicator and fixing the number of offloading vehicles, the results obtained may not have much practical significance. On the other hand, the solution can be found via exhaustive search but it may not be scalable for a large vehicular network due to the complexity of the problem. 

\section{Proposed Solution for Mining Cluster Selection}

In this section, we make use of game-theoretic strategies and bipartite graphs to model the interplay of offloading vehicle and mining cluster and finding the efficient offloading solution. 

We first use the graph theory approach to model different dynamics between offloading vehicles and mining cluster Fig. \ref{fig2} shows. Specifically, to account for different execution tasks of offloading vehicles, the vehicles within a mining cluster must be connected to the offloading vehicles with different weights. Assuming the computation resources within a mining cluster are divided equally among the number of vehicles in a mining cluster\footnote{\textcolor{black}{Note that we have considered commonly used uniform distribution of computation resources for the proposed solution. However, in practical conditions, it is possible that computation resources are distributed non-uniformly depending on the different types of vehicles. This case will be addressed in future studies.}}, we have: 
\begin{align}
R_{i,j}=\frac{1}{S_j},
\end{align}
where $S_j$ denotes the number of offloading vehicles connected to the $j$-th mining cluster. The maximum utility sum of the mining cluster can be given as:
\begin{align}
\sum_{s=1}^{S_j}{\log \biggl(\frac{\Omega_{i,j}}{S_j}\biggr)}=\log 
\biggl(\frac{\sum_{s=1}^{S_j}{\Omega_{i,j}}}{S_j^{S_j}}\biggr)
\end{align}

Using the maximal weighted approach, it can be shown that the sum of weights for the edges of the graph are lower bounded by: 
\begin{align}
\log \left(\frac{\sum_{s=1}^{S_j}{\Omega_{i,j}}}{S_j^{S_j}}\right)\le 
\sum_{s=1}^{S_j}{\log \left\{ \frac{\Omega_{i,j}(s-1)^{s-1}}{s^{s}}\right\} }
\end{align}

Thus, for a given offloading vehicle set ${\mathcal{G}} $ and mining cluster $\mathcal{F} $, a bipartite graph can be constructed. We first introduce the vehicles of a mining cluster as $v_{j}^{1},v_{j}^{2},\ldots ,v_{j}^{V}$, where $j \in \mathcal{F} $. When an $i$-th offloading vehicle is in range of the $j$-th mining cluster, we add an edge between $(u_{i},v_{j}^{s})$ such that $S_j \leq V$, where $s=1,2,\ldots ,V$ with the weight given as:
\begin{align}
w_{i,j}^{s}=\log \left\{ \frac{\Omega_{i,j}\Xi}{s^{s}}\right\}. 
\end{align}

where $\Xi = (s-1)^{s-1}$. When a vehicle offloads a task to the mining cluster, the corresponding mining vehicle $v_{1}^{1}$ would be matched with the weight $\log \Omega_{11}$. When another offloading vehicle offloads a task to the same mining cluster, the cluster head divides the computation resources equally. Due to this reason, the net utility of both offloading vehicles become $\log \left\{\frac{\Omega_{1,1}}{2}\right\}$ and $\log \left\{\frac{\Omega_{2,1}}{2}\right\}$, respectively. The marginal utility gain for adding another vehicle becomes $\log \left\{\frac{\Omega_{2,1}}{4}\right\}=\log \left\{\frac{\Omega_{1,1}}{2}\right\}-\log \left\{\Omega_{1,1}\right\}+\log \left\{\frac{\Omega_{2,1}}{2}\right\}$.

Note that this gain would be equal to the weight between $u_{2}$ and $v_{1}^{2}$. Thus, by offloading the $s$-th vehicle to the mining cluster, we can keep track of the marginal utility gains for a particular mining cluster. \textcolor{black}{It is worth mentioning that there are two parts of the edge weights $w_{i,j}^{s}$.} The first part $\log \Omega_{i,j}$ deals with the offloading rate of the vehicle, while the other part represents the computation burden on the mining cluster. The two parts can be dealt with separately at the mining cluster using an auction-based game theory approach. 
\begin{algorithm}[t]
1:\ \ {\bf Initialization:} Environment parameters and assign $p_j^{s}= -\log\{\frac{\Xi}{s^s}\}$\\
{\bf while} \ {Change in assignment }\\{
2:\ \     $p_j = \min_s{p_j^{s}}$ \; \\
3:\ \     Announce $p_j$ in the network and $s^{*} = \arg{\min_s{p_j^{s}}}$ \; \\
4:\ \     Collect bid from the nearby vehicles \; \\
5:\ \     Select vehicle with maximum bid ($b$) such that ${vehicle}^{*} = \arg{\max_i{b_i}}$ \; \\
6:\ \     Assign temporarily ${vehicle}^{*}$ to $s^{*}$ \; \\
7:\ \     Remove previously assigned ${vehicle}^{*}$ \; \\
8:\ \     Announce temporary assignment in network \; \\}
{\bf end while} \\
\caption{Mining cluster price determination and assignment.}
\end{algorithm}

Initially, the mining cluster generates a price based on the available computation resources given by:
\begin{align}
p_{j}^{s}=-\log\biggl\{\frac{\Xi}{s^{s}}\biggr\}.
\end{align}

The utility of the offloading vehicle is based on the price of each mining cluster. Specifically, every offloading vehicle compares the price of mining clusters and sets its utility as:
\begin{align}
\mathcal{U}_{i,j}=c+\log \Omega_{i,j},
\end{align}
where $c$ is a pre-defined constant with a value greater than the initial prices by mining clusters. The auction process takes place in multiple rounds. A mining cluster announces the price $p_{j}^{s}$ to the other vehicles in the network. Based on the broadcasted price, the offloading vehicle calculates the gain as:
\begin{align}
m_{i,j}=\mathcal{U}_{i,j}-p_{j}^{s}
\end{align}

The offloading vehicle calculates two margins, i.e., $m_{i}^{*}$ and $\tilde{m}_i$ called the highest and second-highest margin respectively. The difference between the two values is submitted as a bid to the mining cluster that provides the largest margin. If there is a tie among two mining clusters, the offloading vehicle submits a bid of $\delta$. To maximize its utility, each mining cluster selects a vehicle with the highest bid and temporarily assigns it to the cluster. In the new round of auction, the mining cluster increases the price by the amount of the highest bid. If a new offloading vehicle provides a higher bid, the temporarily assigned vehicle is removed and a new assignment is performed. The new temporarily assigned vehicle along with the latest price is announced in the network via a broadcast message from the mining cluster. 

\textcolor{black}{The process of auction continues until all the mining clusters stop changing the temporary assignment of the offloading vehicles.} In our version of the auction, all the unassigned vehicles submit their bid simultaneously in a single round. It is also straightforward to extend the approach where vehicles are allowed to submit their bids asynchronously. It may also be noted that the price of the mining clusters increases at least by $\delta$. For a given limit ($\alpha$) on the number of computing vehicles that a mining cluster can have, the maximum number of required rounds for completing the bidding process is bounded by: 

\begin{algorithm}[t]
1:\ \  Evaluate the transmission rate $\Omega_{i,j}$ for all connected mining clusters. \\
{\bf while} \ {Change in assignment }\\{
2:\ \   Collect $p_j$ from connected mining clusters \; \\
3:\ \     Collect assignment to mining clusters \; \\
\ \   {\bf if} \ {Temporary assignment}\\{
4:\ \     Calculate the margins for all nearby mining clusters $m_{i,j} = \mathcal{U}_{i,j}-p_j$ \; \\
5:\ \     Select mining cluster with largest margin $j^{*} = \arg{\max_j{m_{i,j}}}$ \; \\
6:\ \     Calculate the largest $m_i^* = \max_j{m_{i,j}}$ and second largest margins $\tilde{m}_i$. \; \\
\ \ \ \ {\bf if} \ {$m_i^{*} - \tilde{m}_i >0$}\\{
7:\ \ \ \ Submit new bid to selected mining cluster $b_i = m_i^{*} - \tilde{m}_i$ \; \\}
\ \ \ \ {\bf else} \\
    {
8:\ \ \ \ Submit $b_i=\delta$ to $j^*$ \; \\
    }
\ \ \ \    {\bf end If} \\}
\ \    {\bf end If} \\
  }{\bf end while}
    \caption{Offloading vehicle bidding process.}
\end{algorithm}
\begin{align}
\max \lbrace \mathcal{U}_{i,j} \rbrace \times \frac{\alpha }{\delta }.
\label{eq_5}
\end{align}

From (\ref{eq_5}), we note that the number of rounds depends on the computation capabilities of the mining cluster and its overall utility.

\section{Performance Evaluation}

This section presents the simulation results and relevant discussions. \textcolor{black}{As indicated in Section II, we consider the wireless channel between the offloading vehicle and mining clusters to be Rayleigh faded. Rayleigh fading is an important ionospheric and tropospheric radio channel model when there are different signal reflection points in the environment. This type of propagation channel adequately suits the cellular V2X network considered in this work.} \textcolor{black}{We have compared the result of our proposed technique with the baseline nearest cluster selection approach. In this approach, each vehicle selects the cluster based on the smallest Euclidean distance. Nearest cluster selection is a commonly used technique for offloading tasks \cite{xu2018survey}.} 

\begin{figure}[!htp]
\centering
\begin{tabular}{c}
\includegraphics[trim={0 0cm 0 0cm},clip,scale=.32]{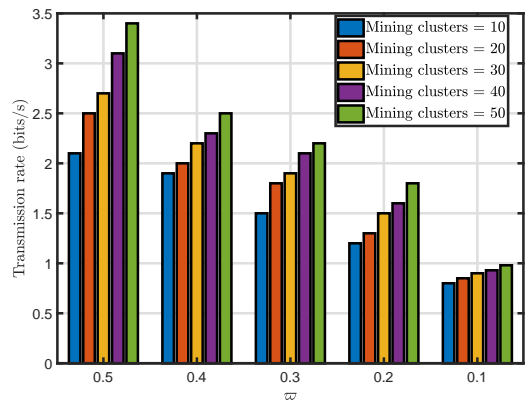}
\end{tabular}
\caption{Transmission rate as a function of $\varpi$, where the transmit power is 25 dBm.}
\label{fig_1_2_1}
\end{figure}

To further demonstrate the impact of using a short packet on the transmission rate, Fig. \ref{fig_1_2_1} shows the results for different values of $\varpi$. Additionally, an increase in the number of mining clusters also improves the transmission rate. However, at smaller values of $\varpi$, the value of the transmission rate decreases significantly. When $\varpi=0.5$, we can observe that the overall transmission rate becomes lower than 1 bit/s. This also decreases the number of mining clusters because there is a negligible change in the transmission rate when the number of mining clusters increases from 10 to 50.  

\begin{figure}[!htp]
\centering
\begin{tabular}{c}
\includegraphics[trim={0 0cm 0 0cm},clip,scale=.32]{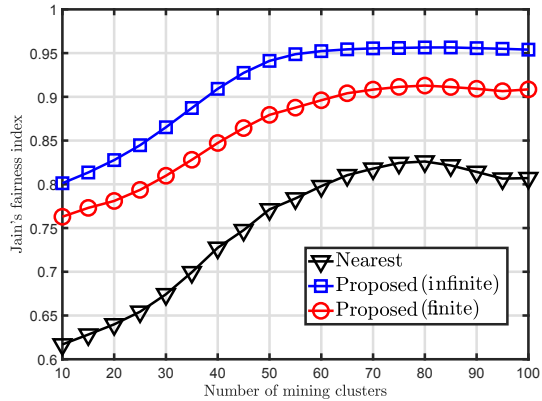}
\end{tabular}
\caption{Jain's fairness index against number of mining clusters.}
\label{fig_2_1}
\end{figure}

To illustrate the benefit of the proposed technique, Fig. \ref{fig_2_1} shows the results for Jain's fairness index of offloading vehicles. We observe that the proposed technique has a higher fairness index as compared to the conventional baseline approach. Specifically, for infinite blocklength, the fairness index is higher compared with the shorter blocklength. However, both alternatives to the proposed technique achieve better fairness index than the nearest cluster selection approach. The major reason for the poor performance of the nearest mining cluster approach is because the offloading vehicles closer to the mining cluster can offload more data which quickly results in saturation in the cluster. As a result, there is an imbalance in the use of network resources due to the selection of the nearest mining cluster. In contrast, the proposed approach does not rely on the proximity and creates a balance in the use of network resources to achieve better fairness and maintain a reasonable tradeoff.

\begin{figure*}[!htp]
\centering
\begin{tabular}{ccc}
\includegraphics[trim={0 0cm 0 0cm},clip,scale=.22]{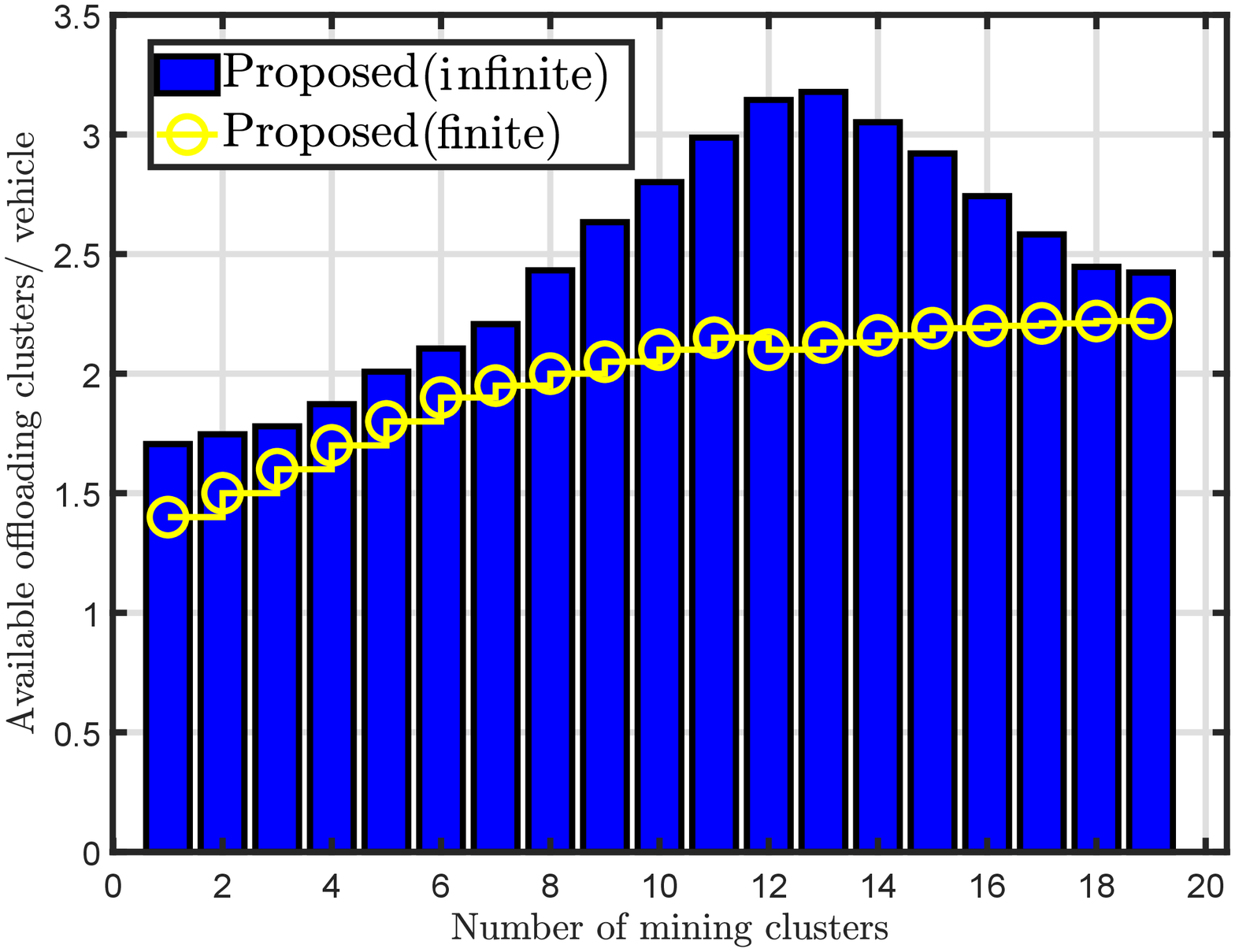} &
\includegraphics[trim={0 0cm 0 0cm},clip,scale=.22]{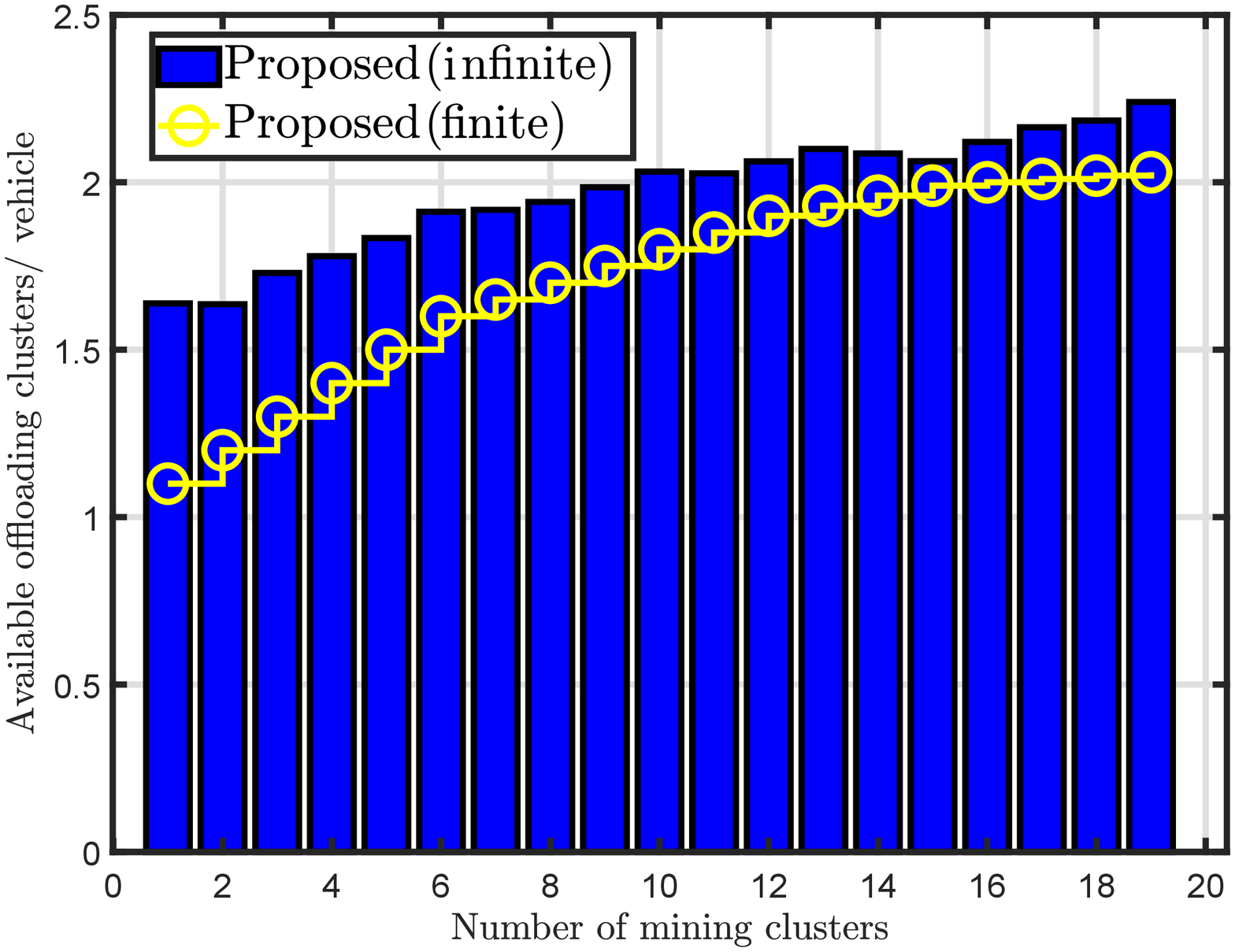} &
\includegraphics[trim={0 0cm 0 0cm},clip,scale=.22]{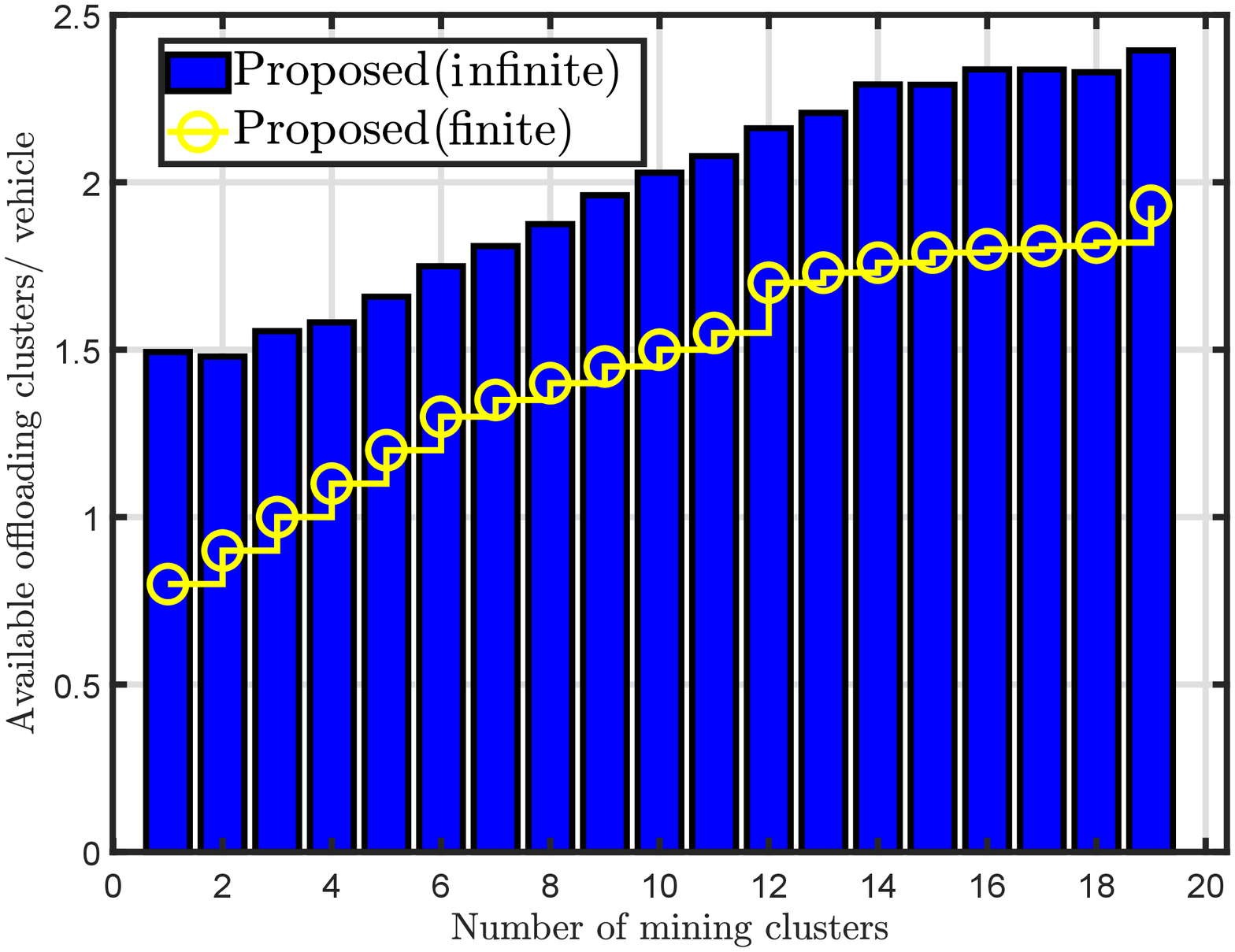} \\
(a) & (b) & (c)
\end{tabular}
\caption{Available mining clusters per vehicle versus number of mining cluster, where (a) $\delta = 10^{-5}$, (b) $\delta = 10^{-4}$, (c) $\delta = 10^{-3}$.}
\label{fig_2_5}
\end{figure*}

Fig. \ref{fig_2_5} shows the average number of available mining clusters per vehicle for different values of $\delta$. We observe that increasing the value of $\delta$ has a direct impact on the cost of the mining cluster. Specifically, during each round, the price increases by $\delta$. The value of $\delta$ changes from $10^{-5}$ to $10^{-3}$ in Fig. \ref{fig_2_5} (a) to (c). Due to this reason, the number of available offloading clusters per vehicle gradually decreases as $\delta$ approaches $10^{-3}$. On the other hand, an increase in each round helps the mining clusters to improve their gains. Besides, we again note that the overall transmission rate decreases for the short blocklength transmission as compared with the infinite blocklength transmission.

\begin{figure}[!htp]
\centering
\begin{tabular}{c}
\includegraphics[trim={0 0cm 0 0cm},clip,scale=.32]{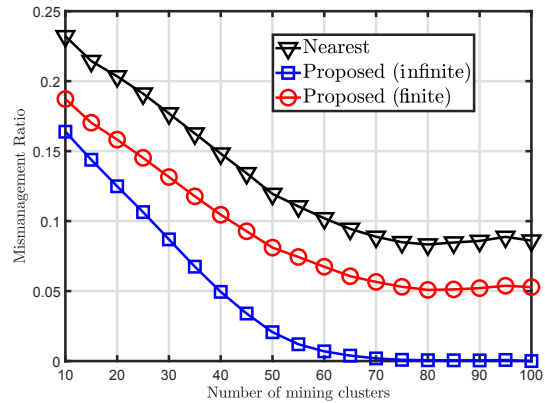}
\end{tabular}
\caption{Mismanagement ratio versus number of mining clusters.}
\label{fig_2_2}
\end{figure}

Fig. \ref{fig_2_2} shows the mismanagement ratio, i.e., the ratio of a vehicle unable to offload to the total number of offloading vehicles in the network. From the figure, we can observe that the mismanagement ratio decreases with an increase in the number of mining clusters. This is intuitive because a larger number of mining clusters allow a better opportunity for offloading vehicles to offload their data. \textcolor{black}{Again, it can be seen that the proposed technique outperforms the baseline method by reducing the number of vehicles that cannot be offloaded.} The difference between the performances of the two techniques becomes clearer when the number of mining clusters increases in the network. 
 
\section{Conclusion and Future Work}

The wide range of applications of blockchain will continue to improve the performance of cellular V2X networks. However, to reap the full benefits of this integration, it is important to address some challenges and identify potential solutions. In this work, we propose an efficient solution for offloading mining tasks in cellular V2X networks. To satisfy the low-latency requirements of safety applications, we have considered a short blocklength transmission architecture. The finite blocklength architecture is a more practical approach to model such blockchain networks. Subsequently, we adopted a game-theoretic approach to efficiently offload the mining tasks to the mining clusters. Our proposed solution not only ensures good transmission rates but also maintains fairness among offloading vehicles. When compared with the conventional baseline nearest cluster selection approach, the results obtained reveal that the performance of the proposed technique improves when the number of mining clusters increases in the network. 

Although the proposed solution provides considerable performance gains, it can be improved in several ways. \textcolor{black}{For instance, the scalability of the blockchain network is still a challenging problem. We anticipate that the cooperation among different mining clusters can be helpful to address the issue of scalability in such networks. Additionally, future studies can jointly consider the mining task offloading and security in blockchain-based cellular V2X networks. Since security is one of the critical issues in conventional blockchain networks, it would be more useful to explore its impact on the offloading performance of vehicular networks.} These challenging yet interesting extensions will be part of our future work. 

\section*{Acknowledgment}

The work of F. Jameel and R. J\"antti was partly supported by Business Finland under the project 5G Finnish Open Research Collaboration Ecosystem (5G-FORCE).

\ifCLASSOPTIONcaptionsoff
  \newpage
\fi

\bibliographystyle{IEEEtran}
\bibliography{Furqan_EE}

\begin{IEEEbiography}[{\includegraphics[width=1in,height=1.25in,clip,keepaspectratio]{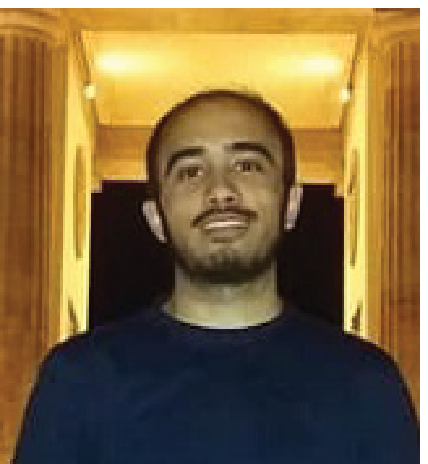}}]{Furqan Jameel} received his B.S. in Electrical Engineering (under ICT R\&D funded Program) in 2013 from the Lahore Campus of COMSATS Institute of Information Technology (CIIT), Pakistan. In 2017, he received his master's degree in Electrical Engineering (funded by prestigious Higher Education Commission Scholarship) at the Islamabad Campus of CIIT. In September 2018, he visited Simula Research Laboratory and the University of Oslo, Norway. From 2018 to 2019, he was with the University of Jyv\"askyl\"a, Finland, and Nokia Bell Labs, Espoo, where he worked as a researcher and a summer trainee, respectively. Currently, he is with the Department of Communications and Networking, Aalto University, Finland, where his research interests include modeling and performance enhancement of vehicular networks, machine/ deep learning, ambient backscatter communications, and wireless power transfer. He is the recipient of outstanding reviewer award 2017 from Elsevier.
\end{IEEEbiography}

\begin{IEEEbiography}[{\includegraphics[width=1in,height=1.25in,clip,keepaspectratio]{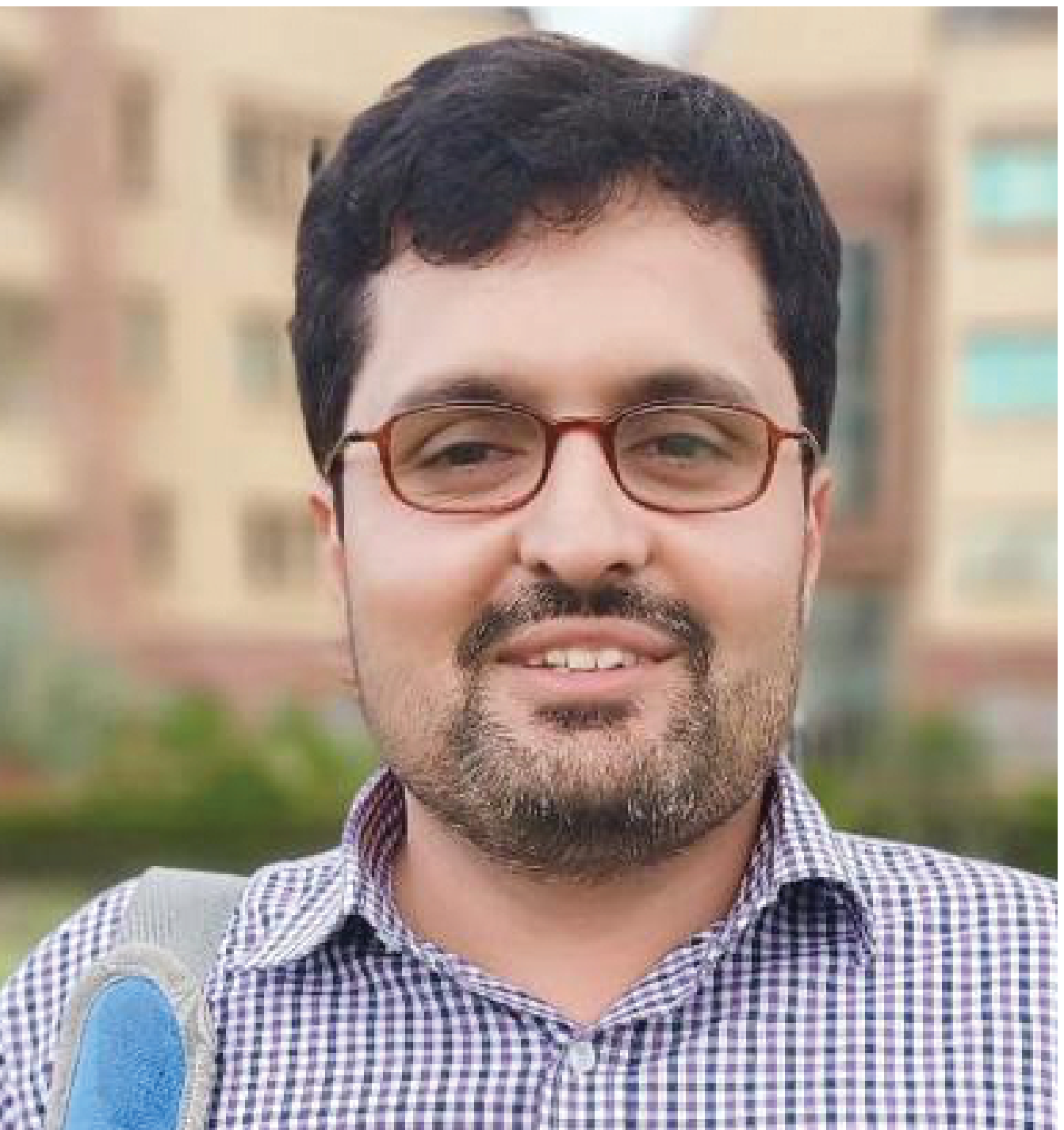}}]{Muhammad Awais Javed} (S'13, M'19, SM'19) is currently working as an Assistant Professor at COMSATS University Islamabad, Pakistan. He completed his Ph.D. in Electrical Engineering from The University of Newcastle, Australia in Feb. 2015 and B.Sc. in Electrical Engineering from University of Engineering and Technology Lahore, Pakistan in Aug. 2008. From July 2015-June 2016, he worked as a Postdoc Research Scientist at Qatar Mobility Innovations Center (QMIC) on SafeITS project. His research interests include intelligent transport systems, vehicular networks, protocol design for emerging wireless technologies and Internet of things.
\end{IEEEbiography}

\begin{IEEEbiography}[{\includegraphics[width=1in,height=1.25in,clip,keepaspectratio]{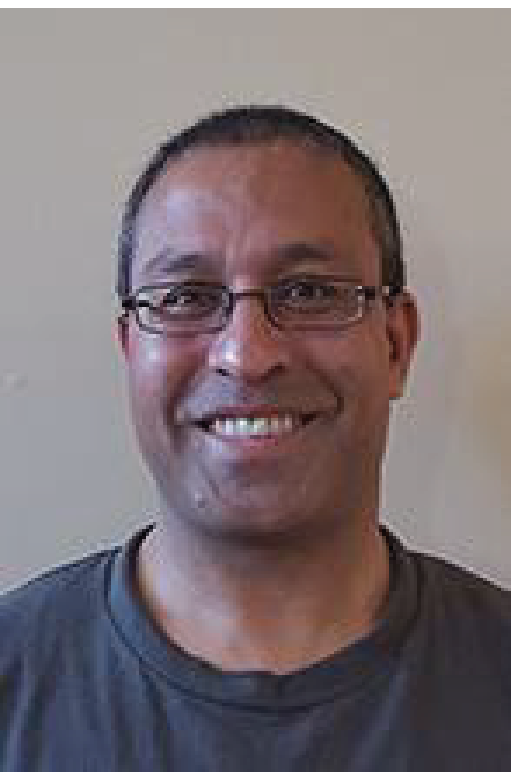}}]{Sherali Zeadally} earned his bachelor's degree in computer science from the University of Cambridge, England.
He also received a doctoral degree in computer science from the University of Buckingham, England, followed by postdoctoral research at the University of Southern California, Los Angeles, CA. He is currently an Associate Professor in the College of Communication and Information, University of Kentucky. His research interests include Cybersecurity, privacy, Internet of Things, computer networks, and energy-efficient networking. He is a Fellow of the British Computer Society and the Institution of Engineering Technology, England.
\end{IEEEbiography}

\begin{IEEEbiography}[{\includegraphics[width=1in,height=1.25in,clip,keepaspectratio]{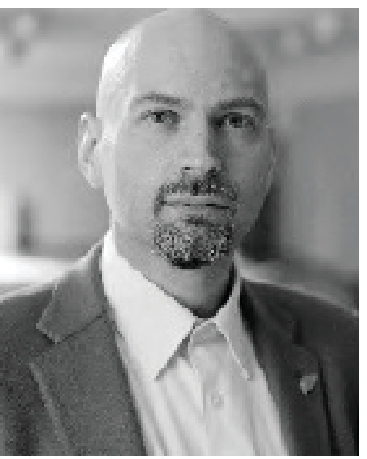}}]{Riku J\"antti} received the M.Sc. degree (Hons.) in electrical engineering and the D.Sc. degree (Hons.) in automation and systems technology from the Helsinki University of Technology (TKK), in 1997 and 2001, respectively. He was a Professor pro term with the Department of Computer Science, University of Vaasa. In 2006, he joined the School of Electrical Engineering, Aalto University (formerly known as TKK), Finland, where he is currently a Professor in communications engineering and the Head of the Department of Communications and Networking. His research interests include radio resource control and optimization for machine type communications, cloud-based radio access networks, spectrum and co-existence management, and RF inference. He is an Associate Editor of the IEEE TRANSACTIONS ON VEHICULAR TECHNOLOGY. He is also an IEEE VTS Distinguished Lecturer (Class 2016).
\end{IEEEbiography}

\end{document}